\documentclass[prl, aps, 12pt, showpacs,epsfig]{revtex4}
\usepackage{graphicx}
\begin{document}

\title{ Leptonic anomalous gauge couplings detection on electron positron colliders  }

\author{Sheng-Zhi Zhao, Bin Zhang}
\email{zb@mail.tsinghua.edu.cn(Communication author)}
 \affiliation{ Department of Physics,
Tsinghua University, Beijing, 100084, China} \affiliation{Center for
High Energy Physics, Tsinghua University, Beijing, 100084, China}

\begin{abstract}
     We studied the dimension-6 leptonic anomalous gauge couplings in the formulation
of linearly realized gauge symmetry effective Lagrangian and investigated the constraints on these anomalous couplings from the existed experimental data including LEP2 and $W$/$Z$ boson decay. Some bounds of $O(0.1-10){\rm TeV}^{-2}$ on four relevant anomalous couplings are given by the Z factories.  We studied the sensitivity of testing the leptonic anomalous couplings via the process $e^+e^-\rightarrow W^+W^-$ at future $e^+e^-$ linear colliders. We discussed different sensitivities to anomalous couplings at polarized and unpolarized $e^+e^-$ colliders, respectively, with 500 GeV and 1 TeV collision energy. Our results show that the a 500 GeV ILC can provide a test of the anomalous couplings, with the same relative uncertainty of cross section measurement, of  $O(0.1-1){\rm TeV}^{-2}$, and a 1 TeV ILC can test the anomalous couplings of  $O(0.01-0.1){\rm TeV}^{-2}$.

\end{abstract}


 \maketitle

\section{Introduction}
   The effective Lagrangian has long been an important model independent approach for studying new physics beyond standard model(SM).
 It is customary to formulate new physics effects by linearly realizing the gauge symmetry
\cite{Leung, Rujula}. After integrating out heavy degrees of freedom
above the cutoff scale, the leading effects at low energies can be
parameterized by the effective interactions. Reference\cite{Leung}
has systemically given all the the gauge-invariant dimension-6
operators which can be constructed from the standard model fields. The
coefficients of these anomalous operators, which are called ``anomalous
couplings'',  reflect the strength of the new physics effects at low
energies. There are already many theoretical studies which have
suggested how to test the anomalous gauge couplings of the Higgs boson
and gauge bosons in the literature for the LHC
\cite{Eboli,Campos,Zeppenfeld,He,Zhang}, and for the ILC
\cite{Barger,Han,Gonzalea}. However,there has been much less effort put into understanding how to detect the anomalous gauge couplings of the fermions at
colliders.

 Since the discovery of neutrino oscillations in recent years\cite{neuosc},
leptonic flavor physics, in particular, the neutrino mass
and flavor mixing has become a hot topic in particle physics. Many theoretical models introduce heavy neutrinos or fourth generation of
leptons to explain the small neutrino masses\cite{seesaw}. In these
models, the small neutrino masses are given by the seesaw
mechanism. The seesaw mechanism generally requires heavy neutrino
masses or very massive fourth generation leptons. It is very difficult
to detect these particles directly at the LHC or other future colliders.
However, the effects of those extra massive particles can be
reflected in the anomalous couplings of leptons and gauge bosons in
the low-energy effective Lagrangian. The measurement of the
phenomenological effects of these leptonic anomalous gauge couplings
on colliders will be useful in understanding the new physics
beyond the standard model related with leptons and neutrinos. Furthermore, many new physical models of electroweak spontaneous symmetry breaking, such as little Higgs models\cite{littleHiggs}, Higgsless models\cite{Higgsless} and Left-Right symmetric gauge models\cite{LRModels}, introduce mixing between extra gauge bosons and the ordinary gauge bosons of the standard model. As a result, there will be anomalous couplings, different from those of the standard model, between the fermions and the ordinary gauge bosons in those models. Detecting these leptonic anomalous gauge couplings on colliders can also test and verify these electroweak new physical models.

In this paper, we proposed the process $e^+e^- \rightarrow W^+W^-$ at
the future electron-positron linear collider (ILC)  to detect the anomalous couplings between the electron and
the $W$, $Z$ gauge bosons. This process is the simplest process at  $e^+e^- $ colliders,
 and both $W$ and $Z$ gauge couplings are involve in the process $e^+e^- \rightarrow W^+W^-$.

In the standard model the $e^+e^- \rightarrow W^+W^-$ process, there is a cancelation between the $E^2$ terms of different Feynman diagrams, therefore the total amplitude and cross
section do not increase with collision energy. If anomalous couplings exist, the cancelation of energy power will be
destroyed. Furthermore, the anomalous couplings will result in higher
energy power dependence. Therefore, the existing low energy electron-positron
experiments give a very weak  limit on the anomalous couplings,
while the anomalous couplings are more likely to be detected at the future
high-energy colliders.

\section{The leptonic anomalous gauge couplings in effective
Lagrangian}
To extend the structure of the SM in a model-independent approach, it is customary to formulate
new physics effects by linearly realizing the gauge symmetry. After integrating out heavy degrees
of freedom above the high scale $\Lambda$, the leading effects at low energies can be parameterized by the effective
interactions

\begin{equation}
\nonumber {\cal L}_{\mbox{eff}} ~\,=~\, \sum_n
\frac{f_n}{\Lambda^2} {\cal O}_n \,,
\end{equation}
where $f_n$'s are dimensionless ``anomalous couplings'', and ${\cal O}_n$'s are the gauge-invariant dimension-6 operators,
constructed from the SM fields.
C. N. Leung, S. T. Love and S. Rao \cite{Leung} described all the dimension-6 $SU_c(3)\times SU_W(2)\times U(1)$
gauge invariant operators. Of all these operators, there are six which involve leptonic gauge couplings and are CP even. They are:

\begin{eqnarray}
&& {\cal O}^{VF}_{7} = i\overline{L}\gamma_\mu W^{\mu\nu} \tensor{D}_\nu L\,,  \nonumber \\
&& {\cal O}^{VF}_{11} = i\overline{L}\gamma_\mu B^{\mu\nu} \tensor{D}_\nu L\,, \nonumber \\
&& {\cal O}^{VF}_{13} = i\overline{E}\gamma_\mu B^{\mu\nu} \tensor{D}_\nu E\,, \nonumber \\
&& {\cal O}^{VF}_{24} = \overline{L}\gamma_\mu (D_\nu W^{\mu\nu}) L\,, \nonumber \\
&& {\cal O}^{VF}_{26} = \overline{L}\gamma_\mu \partial_\nu B^{\mu\nu} L\,, \nonumber \\
&& {\cal O}^{VF}_{27} = \overline{E}\gamma_\mu \partial_\nu B^{\mu\nu}
E\, , \nonumber
\end{eqnarray}

where $L$ is left-hand $SU_W(2)$ lepton doublet and $E$ is lepton right-hand singlet.

The Feynman rules for the vertices involving leptons and gauge bosons are listed in the appendix. From these Feynman rules, we find that the operators
 $ {\cal O}_7 $ and $ {\cal O}_{24} $ contribute to each of the vertices between leptons and gauge bosons in the process $e^+e^-\rightarrow W^+W^-$. These two operators even provide an additional 4-line $l^+l^-W^+W^-$ vertex. However, other operators ${\cal O}_{11} $ , ${\cal O}_{13} $ , ${\cal O}_{26} $ and ${\cal O}_{27} $ only affect the neutral current vertices. Therefore, the process $e^+e^-\rightarrow W^+W^-$ is much more sensitive to operators  $ {\cal O}_7 $ and $ {\cal O}_{24} $ than others. The two operators ${\cal O}_{11} $ and ${\cal O}_{13}$ are only involved in the initial electron positron fusion vertices in the S channel diagrams, and the three momentums of the  neutral current vertices are parallel. The construction of the two operators' Feynman rules determine that their contributions are zero when all three momentums are parallel as we can see from the Feynman rule (b) in the Appendix.

\section{The constraints from LEP2 and W/Z decay}
The measurement of the total cross section of $ W $ pair production on experiment LEP2\cite{L3} can give some limits on the anomalous coupling constants. However, because the collision energy is just beyond the $W$ pair threshold, the $W$ bosons produced at LEP2 typically have low momentum. The effects of dimension-6 operators are very small. In other words, the constraints from LEP2 are very weak.

In this work, we only calculate the cross section of  $e^+e^-\rightarrow W^+W^-$ at tree level, and figure out the relative deviations from the standard model caused by various anomalous couplings. We make the reasonable assumption that the relative deviation will not be changed significantly by radiative correction and detector simulation.
The LEP2 experimental measurement of the cross section for the process $e^+e^-\rightarrow W^+W^-$ is highly consistent with theoretical calculation. The systematic uncertainty of experimental measurement can conversely give constraints on the anomalous couplings. The relative uncertainty of cross section measurement is about $\pm 2\%$ at LEP2\cite{L3}. Here, we conservatively enlarged the experimental uncertainty and theoretical uncertainty to $\pm 5\%$, any cross section deviations beyond $\pm 5\%$ caused by anomalous couplings can be detected at colliders. We also suppose the same uncertainty for future linear colliders.

In accordance with the experiment, we let the final $W$ bosons decay to fermions and impose the following acceptance cuts on final fermions:
\begin{eqnarray}
|\eta|<3, \;\; p_T> 10 \rm GeV. \nonumber
\end{eqnarray}

\begin{figure}[h]
\center
\includegraphics[width=10cm]{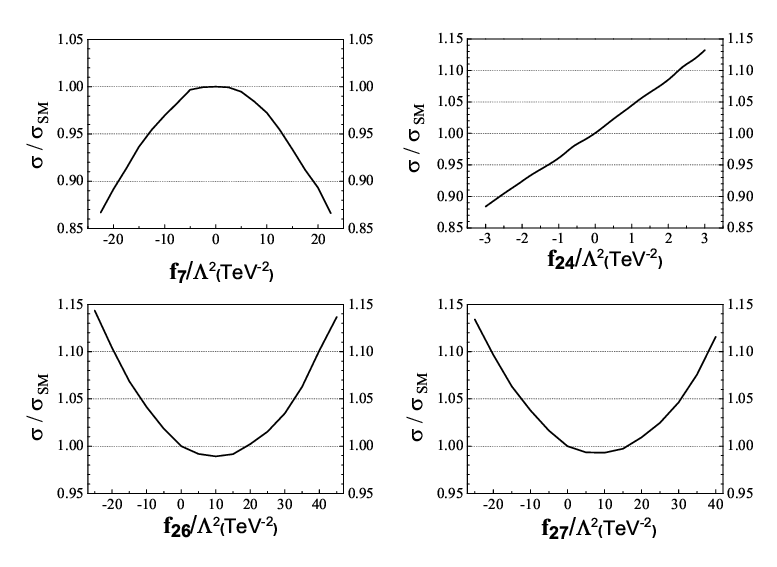}
\caption{The relative deviations of  $e^+e^-\rightarrow W^+W^-$ cross section on LEP2 caused by different anomalous coupling constants. }
\label{fig1}
\end{figure}

After the cuts, the standard model detectable $W$ pair product cross section is $17.7$pb at tree level.  Fig.\ref{fig1} plots the relative deviations caused respectively by different anomalous coupling constants.

According to the experimental relative uncertainty   $\pm 5\%$ , the LEP2 measurement on the cross section can provide the limits on  anomalous couplings:
\begin{eqnarray}
-13 \; {\rm TeV}^{-2}<&f_{7}/\Lambda^2 &<13 \;{\rm TeV}^{-2}, \nonumber \\
-1.1 \; {\rm TeV}^{-2}<&f_{24}/\Lambda^2 &<1.0  \;{\rm TeV}^{-2}, \nonumber \\
-11 \; {\rm TeV}^{-2}<&f_{26}/\Lambda^2 &<33  \;{\rm TeV}^{-2}, \nonumber \\
-13 \; {\rm TeV}^{-2}<&f_{27}/\Lambda^2 &<30 \;{\rm TeV}^{-2}. \nonumber
\end{eqnarray}

The bounds on the four relevant anomalous couplings from LEP2 experiment are very loose, because the outgoing $W$ bosons do not obtain large momentums compared to their mass. However, the dimension-6 anomalous couplings should been enhanced by large momentums.

\begin{figure}[h]
\center
\includegraphics[width=8cm]{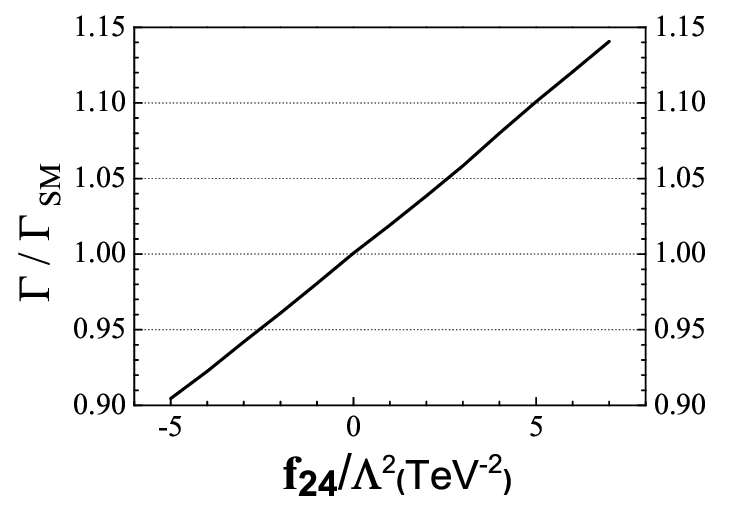}
\caption{The relative deviations of  $W$ boson leptonic decay partial width caused by the anomalous coupling constants $f_{24}$. }
\label{fig2}
\end{figure}

The measurements of the $W$ boson decay width and the leptonic branching ratio have very high accuracy, they also can  provide limits on  anomalous couplings. However only one  anomalous coupling  $f_{24}$ can change the $W$ decay amplitude. In Fig.\ref{fig2} we have plotted the numerical calculated relative deviations of the $W$ boson leptonic decay partial width caused by the anomalous coupling constants $f_{24}$. The $f_{24}$ anomalous coupling can increase or decrease the leptonic decay partial width because of the interference with the standard vertex.

We can also give the analytical form of the the relative deviations of the $W$ boson leptonic decay partial width caused by the anomalous coupling constants $f_{24}$. The $W\ell\nu$ vertex with anomalous coupling is $\frac{i\gamma^{\mu}}{\sqrt{2}}\frac{e}{s} P_L+\frac{i}{\Lambda^2}[k_1{\!\!\!\!\!\slash\;}k_1^\mu-k_1^2\gamma^\mu](-\frac{\sqrt{2}}{2}f_{24})P_L$ as listed in Appendix(c), where $k_1$ is the $W$'s momentum. Because the $W$ boson is on-shell in the decay process, the vertex is simplified to $\frac{i\gamma^{\mu}}{\sqrt{2}}\left(\frac{e}{s}+\frac{M_{W}^{2}}{\Lambda^{2}}f_{24}\right) P_L$. The anomalous coupling $f_{24}$ can change the SM $W$ boson leptonic decay partial width $\Gamma_{sm}\propto (\frac{e}{s})^2 $ to $\Gamma(f_{24})\propto (\frac{e}{s}+\frac{M_{W}^{2}}{\Lambda^{2}}f_{24})^2$. When the anomalous coupling is small, the relative deviations of the $W$ boson leptonic decay partial width becomes:

\begin{eqnarray}
\frac{\Delta \Gamma(f_{24})}{\Gamma_{sm}}\approx \frac{2M_{W}^{2}s}{e}\frac{f_{24}}{\Lambda^{2}} . \nonumber
\end{eqnarray}
It is the same as the numerical calculation result plotted in the Fig.\ref{fig2}.

 The experimental relative uncertainties of the $W$ boson decay width and the leptonic branching ratio are very small, both within $\pm 2\%$\cite{PDG}, so the  relative uncertainty of the leptonic decay partial width is also  within $\pm 2\%$. Therefore, the measurement of the $W$ boson decay  can provide a more stringent limit on  anomalous coupling  $f_{24}$:

\begin{eqnarray}
-1.03 \;{\rm TeV}^{-2}<f_{24}/\Lambda^2 <1.03 \;{\rm TeV}^{-2}. \nonumber
\end{eqnarray}

The measurement of $Z$ boson leptonic partial decay width  has much higher accuracy than that of the $W$ boson decay, so it can  provide more stringent limits on  anomalous couplings. There are three anomalous couplings $f_{24},f_{26}$ and $f_{27}$ which can affect the $Z$ decay amplitude. The three  anomalous couplings change the $Z\ell\ell$ vertex to be $i\gamma^{\mu}\left(\frac{e(2s^{2}-1)}{2cs} P_L+\frac{es}{c}P_R\right)+\frac{i}{\Lambda^2}[k_1{\!\!\!\!\!\slash\;}k_1^\mu-k_1^2\gamma^\mu](\frac{c}{2}f_{24}P_L+sf_{26}P_L+sf_{27}P_R)$ as listed in Appendix(a), where $k_1$ is the $Z$'s momentum. When the $Z$ boson is on-shell in the decay process, the vertex is simplified to
\begin{eqnarray}
i\gamma^{\mu}\left(\frac{e(2s^{2}-1)}{2cs}-\frac{M_{Z}^{2}}{\Lambda^{2}}(\frac{c}{2}f_{24}+sf_{26})\right)P_L+i\gamma^{\mu}\left(\frac{es}{c}-\frac{M_{Z}^{2}}{\Lambda^{2}}sf_{27}\right)P_R.\nonumber
\end{eqnarray}
While the anomalous couplings are small, the relative deviations of the $Z$ boson leptonic decay partial width becomes:
\begin{eqnarray}
\frac{\Delta \Gamma}{\Gamma_{sm}}\approx \frac{\frac{e(2s^{2}-1)}{2s}\frac{ M_Z^2}{\Lambda^{2}}(f_{24}+\frac{s}{2c}f_{26})+2\frac{es}{c}\frac{M_Z^2}{\Lambda^{2}}s f_{27}}{\left(\frac{e(2s^{2}-1)}{2cs}\right)^2+\left(\frac{es}{c}\right)^2}. \nonumber
\end{eqnarray}

 The experimental relative uncertainty of the $Z$ boson  leptonic partial decay width is very small, within $\pm 0.2\%$\cite{PDG}. Therefore, the measurement of the $Z$ boson decay  can provide much more stringent limits on  anomalous couplings:
\begin{eqnarray}
-0.1 \;{\rm TeV}^{-2}<f_{24}/\Lambda^2 <0.1 \;{\rm TeV}^{-2}, \nonumber \\
-0.09 \;{\rm TeV}^{-2}<f_{26}/\Lambda^2 <0.09 \;{\rm TeV}^{-2}, \nonumber \\
-0.11 \;{\rm TeV}^{-2}<f_{27}/\Lambda^2 <0.11 \;{\rm TeV}^{-2}. \nonumber
\end{eqnarray}

 The measurements of the Z effective axial-vector couplings $g_A$ to charged leptons are also the most accurate experiments at the Z factories together with the Z boson decay width. We may also analyze the constraint from the high-precision measurements of $g_A$. Three  anomalous couplings can provide the extra axial-vector coupling, we define the anomalous  axial-vector coupling as $f_A=(f_{24}+\frac{2s}{c} f_{26}-\frac{2s}{c} f_{27})$. The  axial-vector current vertex becomes $  i \gamma^{\mu}\gamma^{5}(-\frac{e}{2cs}g_A+\frac{M_{Z}^2}{\Lambda^2}\frac{c}{4}f_A)$ when Z is on-shell, where $g_A$ is the effective axial-vector coupling of the Z boson to charged leptons with $g_A=0.5$ in standard model. Meanwhile the  high-precision measurements of $g_A$ can give a stringent bound on the anomalous  axial-vector coupling $f_A$. The high-precision $g_A$ magnitude is derived from measurements of the Z line-shape and the forward-backward lepton asymmetries as a function of energy around the Z mass, the measurements have the collision energy close to the Z mass and the Z bosons are almost on-shell. Hence the relative deviation of $g_A$ caused by the axial-vector coupling $f_A$ is
 \begin{eqnarray}
\frac{\Delta g_A}{g_A}=-\frac{M_Z^2 c^2s}{e}\frac{f_A}{\Lambda^2}.
\end{eqnarray}

 The experimental relative uncertainty on  $g_A$ is very small, within $\pm 0.1\%$\cite{PDG}. In order to limit the deviation of $g_A$ to within $\pm 0.1\%$, a very stringent limit on  anomalous couplings combination $f_A$ we have:
  \begin{eqnarray}
-0.1 \;{\rm TeV}^{-2}<f_{A}/\Lambda^2 <0.1 \;{\rm TeV}^{-2}. \nonumber
\end{eqnarray}

\section{The detection of leptonic anomalous gauge couplings at ILC}

\begin{figure}[h]
\center
\includegraphics[width=10cm]{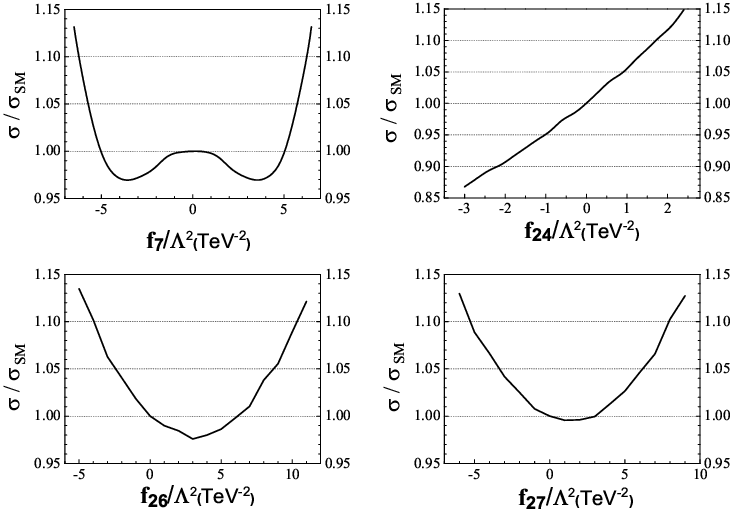}
\caption{The relative deviations of  $e^+e^-\rightarrow W^+W^-$ cross section on 500 GeV ILC caused by various anomalous couplings. }
\label{fig3}
\end{figure}

\begin{figure}[h]
\center
\includegraphics[width=10cm]{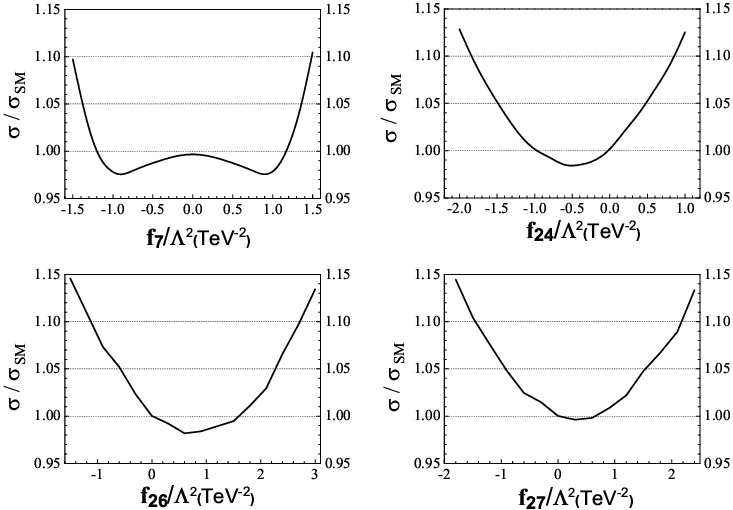}
\caption{The relative deviations of  $e^+e^-\rightarrow W^+W^-$ cross section on 1 TeV ILC caused by various anomalous couplings. }
\label{fig4}
\end{figure}

The anomalous couplings bring  high energy power dependence to the $e^+e^-\rightarrow W^+W^-$ cross section. The higher the collision energy, the greater the deviation caused by anomalous couplings. We have considered the  relative deviations caused by different anomalous couplings at the future ILC with the collision energy $\sqrt{s_{ee}}=500$ GeV in Fig.\ref{fig3}, and at $\sqrt{s_{ee}}=1$ TeV ILC in  Fig.\ref{fig4}. The total cross section decreases as the collision energy increases and is concentrated in the forward region. The quarks or leptons produced in the decays of the energetic $W$ bosons follow the forward alignment such that the basic acceptance cuts reduce the cross section more severely. The standard model total cross section after basic cuts is about $5.68$pb as $\sqrt{s_{ee}}=500$ GeV and $1.42$pb as $\sqrt{s_{ee}}=1$ TeV. However, the large luminosity of ILC will still provide sufficient events to analyze the effect of the anomalous couplings.  We also suppose the ILC can reach  the same experimental relative uncertainty on cross section as LEP2 and the constraint on the detection ability still comes from the experimental uncertainty.

The detection ability of high energy ILC is much better than LEP2. If we also suppose that the same experimental relative uncertainty of cross section on ILC as $\pm 5\%$, the detection sensitivities of the anomalous coupling constants of $\sqrt{s_{ee}}=500$ GeV ILC are:
\begin{eqnarray}
-5.8 \; {\rm TeV}^{-2}<&f_{7}/\Lambda^2 &<5.8 \;{\rm TeV}^{-2}, \nonumber \\
-1.0 \; {\rm TeV}^{-2}<&f_{24}/\Lambda^2 & <0.9  \;{\rm TeV}^{-2}, \nonumber \\
-2.3 \; {\rm TeV}^{-2}< & f_{26}/\Lambda^2 &<8.6  \;{\rm TeV}^{-2}, \nonumber \\
-3.3 \; {\rm TeV}^{-2}< & f_{27}/\Lambda^2 & <6.2 \;{\rm TeV}^{-2}. \nonumber
\end{eqnarray}

And, the detection sensitivities of the anomalous coupling constants of $\sqrt{s_{ee}}=1000$ GeV ILC are:

\begin{eqnarray}
-0.85 \;{\rm TeV}^{-2}<&f_{7}/\Lambda^2& <0.85 \;{\rm TeV}^{-2}, \nonumber \\
-1.5 \;{\rm TeV}^{-2}<&f_{24}/\Lambda^2 & <0.5 \;{\rm TeV}^{-2}, \nonumber \\
-0.6  \; {\rm TeV}^{-2}<&f_{26}/\Lambda^2 &<2.3  \; {\rm TeV}^{-2}, \nonumber \\
-0.9  \; {\rm TeV}^{-2}<&f_{27}/\Lambda^2 &<1.5  \; {\rm TeV}^{-2}. \nonumber
\end{eqnarray}

The detection sensitivities at ILC are increased by about 1 orders of magnitude than at the LEP2.

\section{The $W$ boson angular distribution of the anomalous signature  }

We also analyzed the distribution of the cross section in order to find the sensitive region of the anomalous coupling.
The $W$ outgoing angle is the unique kinematic parameter in the process $e^+e^-\rightarrow W^+W^-$ at the energy determined $e^+e^-$ colliders, and the $W$ angle can be reconstructed from the $W$ decay final states on ILC.
Therefore, we analyzed the  angular distribution of the final state W particle, and compared the distribution differences
between anomalous couplings and the standard model.

\begin{figure}[h]
\center
\includegraphics[width=6cm]{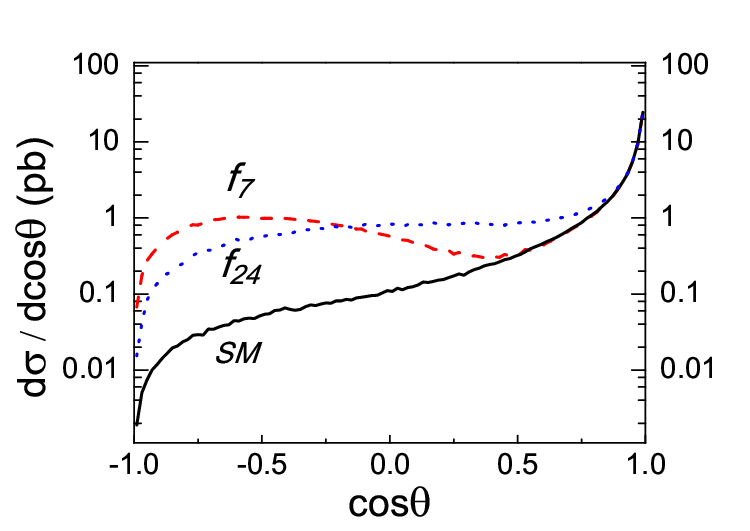}\includegraphics[width=6cm]{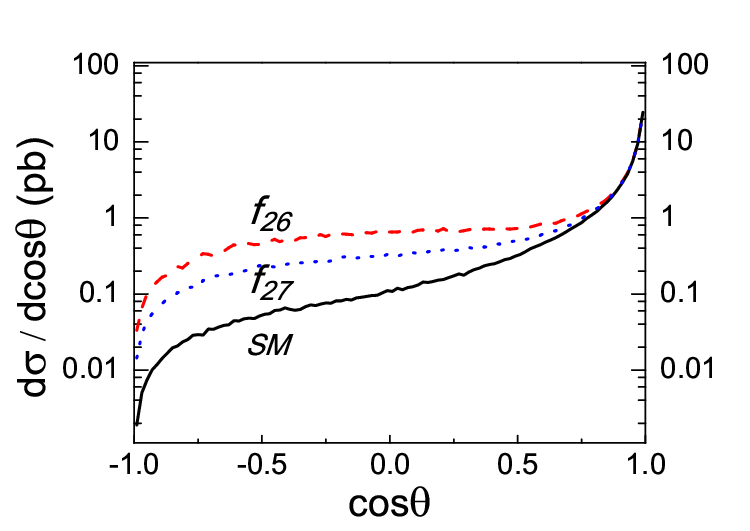}
\caption{The W boson outgoing angle distribution at 1 TeV ILC. The solid curve is the SM distribution, the dashed and dotted curves are the distributions respectively with $f_{7}/\Lambda^2=2\;{\rm TeV}^{-2}$; $f_{24}/\Lambda^2=3\;{\rm TeV}^{-2}$; $f_{26}/\Lambda^2=6\;{\rm TeV}^{-2}$; $f_{27}/\Lambda^2=3\;{\rm TeV}^{-2}$. }
\label{fig_cos1000}
\end{figure}

\begin{figure}[h]
\center
\includegraphics[width=6cm]{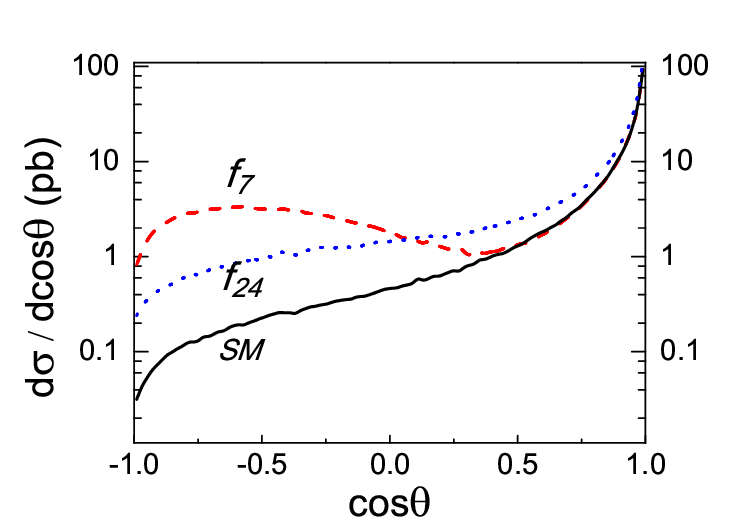}\includegraphics[width=6cm]{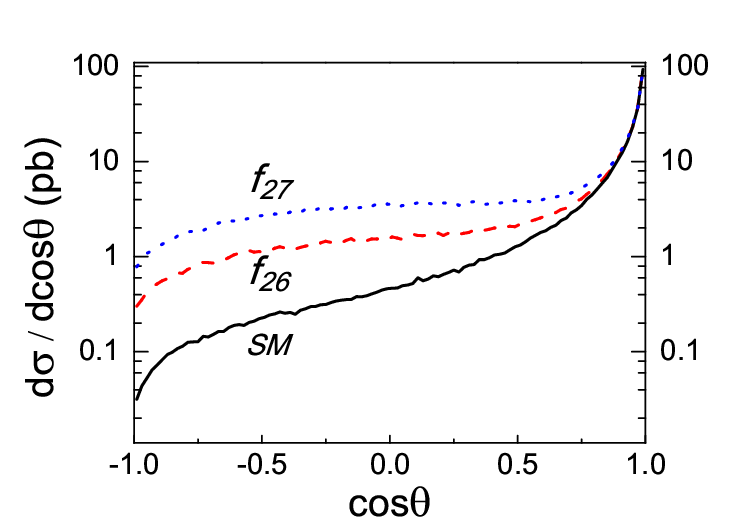}
\caption{The W boson outgoing angle distribution at 500 GeV ILC. The solid curve is the SM distribution, the dashed and dotted curves are the distributions respectively with $f_{7}/\Lambda^2=8\;{\rm TeV}^{-2}$; $f_{24}/\Lambda^2=7\;{\rm TeV}^{-2}$; $f_{26}/\Lambda^2=15\;{\rm TeV}^{-2}$; $f_{27}/\Lambda^2=20\;{\rm TeV}^{-2}$. }
\label{fig_cos500}
\end{figure}

\begin{figure}[h]
\center
\includegraphics[width=6cm]{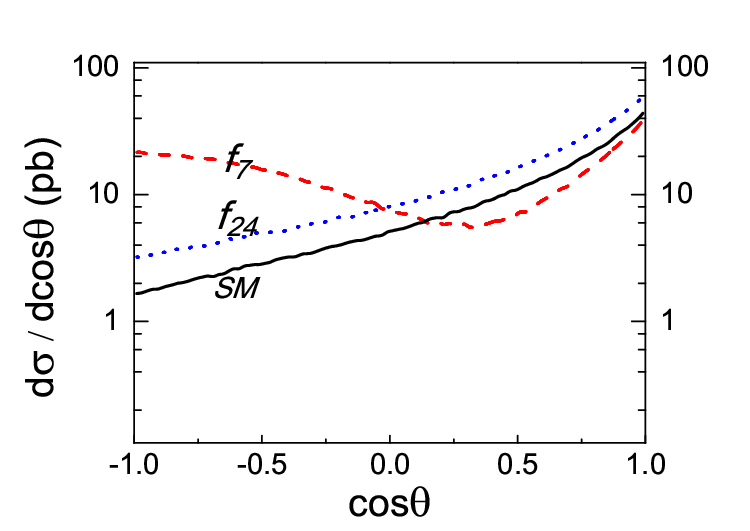}\includegraphics[width=6cm]{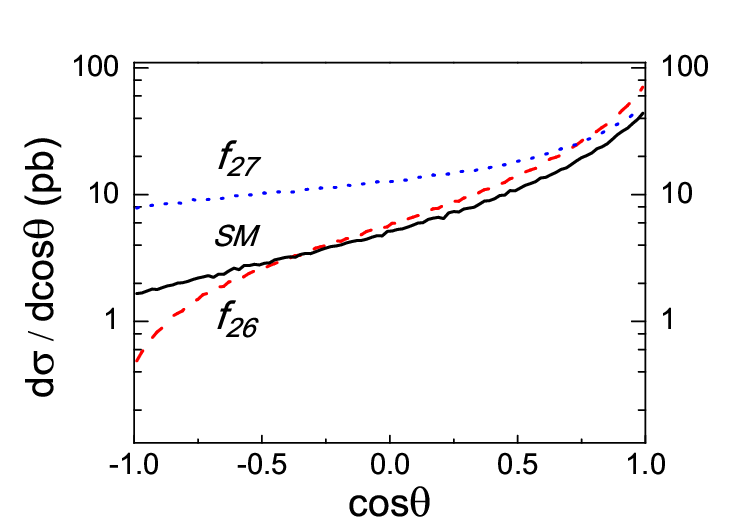}
\caption{The W boson outgoing angle distribution at 200 GeV LEP2. The solid curve is the SM distribution, the dashed and dotted curves are the distributions respectively with $f_{7}/\Lambda^2=63\;{\rm TeV}^{-2}$; $f_{24}/\Lambda^2=10\;{\rm TeV}^{-2}$; $f_{26}/\Lambda^2=60\;{\rm TeV}^{-2}$; $f_{27}/\Lambda^2=90\;{\rm TeV}^{-2}$.}
\label{fig_cos200}
\end{figure}

 Fig.\ref{fig_cos1000} and \ref{fig_cos500} show the differences on $W^-$ boson outgoing angle distribution, where $\theta_{W^-}$ is the $W^-$ production angle with respect to the direction of the incoming electrons. The distribution of the $W^+$ production angle with respect to the direction of the incoming positrons $\theta_{W^+}$ is same for the charge symmetry. In order to give the distribution of the angle of the $W^-$ or $W^+$ bosons, it is necessary to distinguish the sign of the $W$ charge in the experiment. Therefore,  Only the semileptonic decay of the $W$ pairs can be considered, $e^+e^- \rightarrow W^+W^-\rightarrow qq  \ell v$, here the final leptons only include $\mu^\pm$ and $e^\pm$. So the semileptonic decay branching ratio is $66\% \times 21\%\times 2=27.7\%$. For the standard model, the cross section is mainly distributed in the forward region where $\cos\theta_{W^\pm}$ is close to one, because the T channel neutrino exchange diagram is dominant. The anomalous couplings' relative effect to the cross section is much larger in the small  $\cos\theta_{W^\pm}$ area, but the partial cross section in this region is very small as shown in the figures. The trend becomes more obvious with the higher collision energy. The standard model cross section within $\cos\theta_{W^\pm}<0.75$ is $0.28$ pb at a 1 TeV ILC, and decreases to $0.146(0.053)$ pb within $\cos\theta<0.5(0)$. If the integrated luminosity of ILC is large enough (for example, $\int {\cal L} \geq 46.2\; \textrm{fb}^{-1} $), one can apply a $W$ angle cut (for instance $\cos\theta<0.75$) to improve the detection sensitivity. In this case, the standard model will provide more than $\sigma_{SM}(\cos\theta<0.75)\times Br\times \int {\cal L} =3600$ events, and the significance of $5\%$ relative deviation caused by anomalous couplings will be greater than $3\sigma$ (the significance defined as $N_s/\sqrt{N_B}$).  If the integrated luminosity becomes greater, we can take a more stringent $W$ angle cut. The cut $\cos\theta<0.5$ can be applied and still keep  $3\sigma$  significance when $\int {\cal L} \geq 90.2\; \textrm{fb}^{-1} $ and the cut $\cos\theta<0$ can be applied if $\int {\cal L} \geq 245 \;\textrm{fb}^{-1}$ as listed in Table \ref{table_xsection}. The total cross section at a $500$ GeV ILC is larger than at a 1 TeV ILC, and the effect of $W$ outgoing angle cut at a $500$ GeV ILC is not as obvious as at a 1 TeV ILC, so the requirement of the integrated luminosity is much smaller. For the process at the 200 GeV LEP2, as shown in Fig.\ref{fig_cos200}, the outgoing $W$ bosons are not very forward because the $W$ momentum is small compared to its rest mass. The $W$ outgoing angle cut at the $200$ GeV LEP2 is almost not helpful to improve the detection limits of anomalous couplings, except a little improvement on $f_7$.

\begin{table}[h]
\caption{The cross sections after different $W$ angle cuts on $\sqrt{s_{ee}}=500$ GeV and 1 TeV ILC and the required integrated luminosity for 5\% relative deviation to keep $3\sigma$ significance.  }\label{table_xsection}
\begin{tabular}{|c|c|c|c|c|}
 \tableline
$\cos\theta$ cut & $\sigma$ (500 GeV) & required $\int {\cal L}$   & $\sigma$ (1 TeV) & required $\int {\cal L}$   \\

 \tableline
none  & 5.68 pb  & 0.63 $\rm {fb}^{-1}$ & 1.42 pb & 2.5 $\rm {fb}^{-1}$ \\
 \tableline
 $\cos\theta<0.75$ & 1.14 pb$\times 27.7\%$ &  11.2 $\rm {fb}^{-1}$ & 0.28 pb $\times 27.7\%$ & 46.2 $\rm {fb}^{-1}$ \\
  \tableline
$\cos\theta<0.5$ & 0.61 pb$\times 27.7\%$  & 21.3 $\rm {fb}^{-1}$ & 0.146 pb$\times 27.7\%$ &   90.2 $\rm {fb}^{-1}$ \\
  \tableline
  $\cos\theta<0$ & 0.23 pb$\times 27.7\%$  & 56.3 $\rm {fb}^{-1}$ & 0.053 pb$\times 27.7\%$ &  245 $\rm {fb}^{-1}$ \\
  \tableline
 \end{tabular}
\end{table}

The more stringent $W$ angle cut can give the better sensitivity on anomalous couplings detection, while the remaining cross section is lower and required luminosity becomes greater, as listed in Table.\ref{table_xsection}.

\begin{table}[h]
\caption{The detection limits of the anomalous coupling constants on $\sqrt{s_{ee}}=500$ GeV ILC with different $W$ outgoing angle cuts. }\label{table_TL500}
\begin{tabular}{|c|c|c|c|c|}
 \tableline
$\cos\theta$ cut & $f_7/\Lambda^2 (\rm TeV^{-2})$ & $f_{24}/\Lambda^2 (\rm TeV^{-2})$  & $f_{26}/\Lambda^2 (\rm TeV^{-2})$ & $f_{27}/\Lambda^2 (\rm TeV^{-2})$  \\

 \tableline
none  & $-5.8 \sim 5.8$   & $-1.0 \sim 0.9$  & $-2.3 \sim 8.6$  & $-3.3 \sim 6.2$  \\
 \tableline
 $\cos\theta<0.75$ & $-1.5 \sim 1.5$  &  $-0.50 \sim 0.46$  & $-0.64 \sim 0.81$  &$-1.2 \sim 4.0$ \\
  \tableline
$\cos\theta<0.5$ & $-1.1 \sim 1.1$  & $-0.39 \sim 0.33 $ & $-0.41 \sim 0.49$ &  $-0.76 \sim 3.6 $ \\
  \tableline
  $\cos\theta<0$ & $-0.78 \sim 0.78$  & $-0.26 \sim 0.24 $ & $-0.25 \sim 0.28$ &  $-0.49 \sim 0.99 $  \\
  \tableline
 \end{tabular}
\end{table}

\begin{table}[h]
\caption{The detection limits of the anomalous coupling constants on $\sqrt{s_{ee}}=1$ TeV ILC with different $W$ outgoing angle cuts. }\label{table_TL1000}
\begin{tabular}{|c|c|c|c|c|}
 \tableline
$\cos\theta$ cut & $f_7/\Lambda^2 (10^{-1}\rm TeV^{-2})$ & $f_{24}/\Lambda^2 (10^{-1}\rm TeV^{-2})$  & $f_{26}/\Lambda^2 (10^{-1}\rm TeV^{-2})$ & $f_{27}/\Lambda^2 (10^{-1}\rm TeV^{-2})$  \\

 \tableline
none  & $-8.5 \sim 8.5$    & $-15 \sim 5.0$ & $-6 \sim 23 $ & $-9 \sim 15$ \\
 \tableline
 $\cos\theta<0.75$ &  $-3.4 \sim 3.4$   &  $-2.0 \sim 1.8$ & $-1.9 \sim 2.3$   &  $-3.3 \sim 10$ \\
  \tableline
$\cos\theta<0.5$ & $-2.6 \sim 2.6$  & $-1.3 \sim 1.2$ & $-1.1 \sim 1.4  $ &  $-2.0 \sim 9.2 $ \\
  \tableline
  $\cos\theta<0$ & $-1.8 \sim 1.8$  & $-0.8 \sim 0.8$ & $-0.75 \sim 0.78 $ & $-1.4 \sim 2.5 $ \\
  \tableline
 \end{tabular}
\end{table}

Table \ref{table_TL500} and Table \ref{table_TL1000} give the detection limits of the  $\sqrt{s_{ee}}=500$ GeV and  $1$ TeV ILC respectively. Within the limits, the relative deviations of the cross section caused by anomalous couplings are less than $\pm 5\%$ which can not be detected at the ILC. If any listed anomalous couplings go beyond their relevant bounds, the ILC can survey the deviation from the standard model. The more stringent a $W$ angle cut applied, the better detection sensitivity ILC can provide.  We can see from Table \ref{table_TL1000}, after  $W$ angle cut, the ILC  detection sensitivity are increased by 1-2 orders of magnitude than the LEP2.

\section{The polarization scheme}
We also considered ILC with the polarization scheme.
Both in standard model and  anomalous couplings, the $W$ boson only couples to left-handed leptons. If the initial electrons are right handed polarized, they only couple to neutral gauge boson $Z$ and $\gamma$. So the process $e^+_L e^-_R\rightarrow W^+W^-$ only has $Z/\gamma$ S channel Feynman diagrams. Therefore the ILC with polarization scheme will be  very useful to detect the  anomalous coupling between lepton and $Z$ boson. The standard model's right-handed polarized cross section is $107$ fb with  $\sqrt{s_{ee}}=500$ GeV and $22.2$ fb with  $\sqrt{s_{ee}}=1$ TeV. The cross section is still large enough to analysis $\pm5\%$ relative deviation with a certain integrated luminosity.

\begin{figure}[h]
\center
\includegraphics[width=6cm]{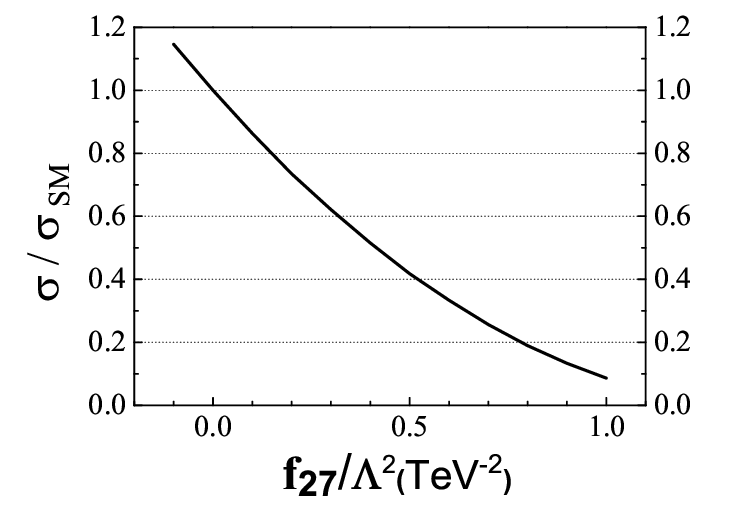}\includegraphics[width=6cm]{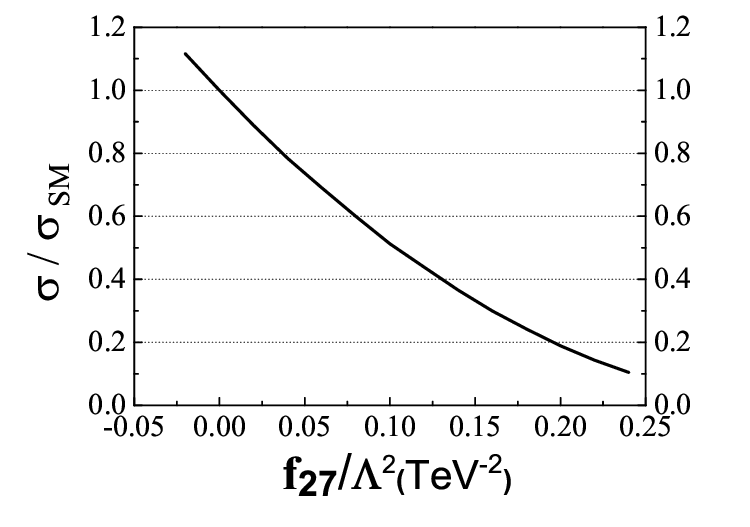}
\caption{The relative deviations of  $e^+_L e^-_R\rightarrow W^+W^-$  cross section at 500 GeV(left) and 1 TeV(right) ILC  and caused by different anomalous coupling constants. }
\label{fig_pol}
\end{figure}

It is different from the unpolarized ILC scheme, the process with right-handed polarized electrons is very sensitive to the coefficient $f_{27}$ which only appears in the anomalous coupling between lepton and $Z$ boson and is not sensitive in the unpolarized $e^+ e^-\rightarrow W^+W^-$ process. In the polarized S channel process, the anomalous coupling $f_{27}$ can interfere with the standard model coupling, and the relative deviations of the cross section are shown in Fig.\ref{fig_pol}.

If we also suppose  the same experimental relative uncertainty of the cross section at the polarized ILC as $\pm 5\%$, the  polarized  $\sqrt{s_{ee}}=500$ GeV ILC can provide a better  detection sensitivity of the anomalous coupling constants  $f_{27}$ :

\begin{eqnarray}
-3.5\times 10^{-2} \;{\rm TeV}^{-2}<f_{27}/\Lambda^2 <3.5\times 10^{-2} \;{\rm TeV}^{-2}, \nonumber
\end{eqnarray}

And for $\sqrt{s_{ee}}=1$ TeV polarized ILC, the detection sensitivity is:
\begin{eqnarray}
-9\times 10^{-3} \;{\rm TeV}^{-2}<f_{27}/\Lambda^2 <9\times 10^{-3} \;{\rm TeV}^{-2}, \nonumber
\end{eqnarray}

Both anomalous coupling  $f_{27}$ and standard model couplings appear in the same S channel Feynman diagrams, so the partial cross section is independent of the $W$ outgoing angle and there is no kinematic difference between the anomalous coupling signal and standard model background. Although we can not find any effective cuts to improve the sensitivity of the right handed polarized electron-positron process, the ability of the polarized linear collider to detect anomalous coupling  $f_{27}$ is still improved by more than one order of magnitude compared to the unpolarized one.

\section{Conclusions}
In this paper, we studied different electron-positron colliders detection abilities on various effective Lagrangian coefficients of LEP2. We analyzed the anomalous couplings' effects on the $e^+ e^-\rightarrow W^+W^-$ cross sections at LEP2,  polarized and unpolarized ILC with 500 and 1000 GeV collision energy.
We only gave the results of the single-parameter study, we simply analyzed the effect of one anomalous coupling and set all others to zero. Single-parameter study is a standard method when considering anomalous couplings measurement.  Of course, different anomalous couplings can affect the same process. While the anomalous couplings are the small deviations from SM couplings, their contributions come from the interference with SM couplings and the interference between different anomalous couplings can be ignored. Therefore, multi-parameter contributions are almost the simple sum of single parameter contributions when the deviations from SM are small. If there is not any deviation found at future experiments, the single parameter study can provide a constraint on anomalous couplings. If there are some deviations found at future experiments, we should study the different kinematics distributions, different polarization scheme or other processes to figure out what anomalous couplings cause the deviations.

Our calculations show that the higher the collision energy, the greater the ability to detect anomalous couplings. The detection sensitivity can be improved by 1-2 orders of magnitude, only when the collision energy increases by several times. This work illustrates as an example that the TeV energy linear colliders have great capability on precision measurement. The future linear colliders have multiple options, such as polarized electron-positron beam and high energy photon colliders. Such options are helpful to measure different vertices.  The scheme with right handed polarized electrons is helpful for detecting the $Z$ boson anomalous couplings. In this paper, we only study the single-parameter effect on the single process  $e^+ e^-\rightarrow W^+W^-$, any anomalous couplings beyond the detection limits can change the cross section significantly. If there are no deviations from the standard model's cross section, then all the four anomalous couplings mentioned above are within the detection bounds. However, once the future measurements observe a cross section deviation, it is difficult to determine which anomalous coupling is responsible. In this case, the study on other processes is useful such as  $e^+ e^-\rightarrow ZZ(Z\gamma)$.

{\bf Acknowledgment:} This work of B.~Z. is supported by the
National Science Foundation of China under Grant No. 11075086 and
11135003.

\section{Appendix}
The Feynman rules of the dimension-6 leptonic anomalous gauge couplings:
\begin{figure}[h]
\center
\includegraphics[width=12cm]{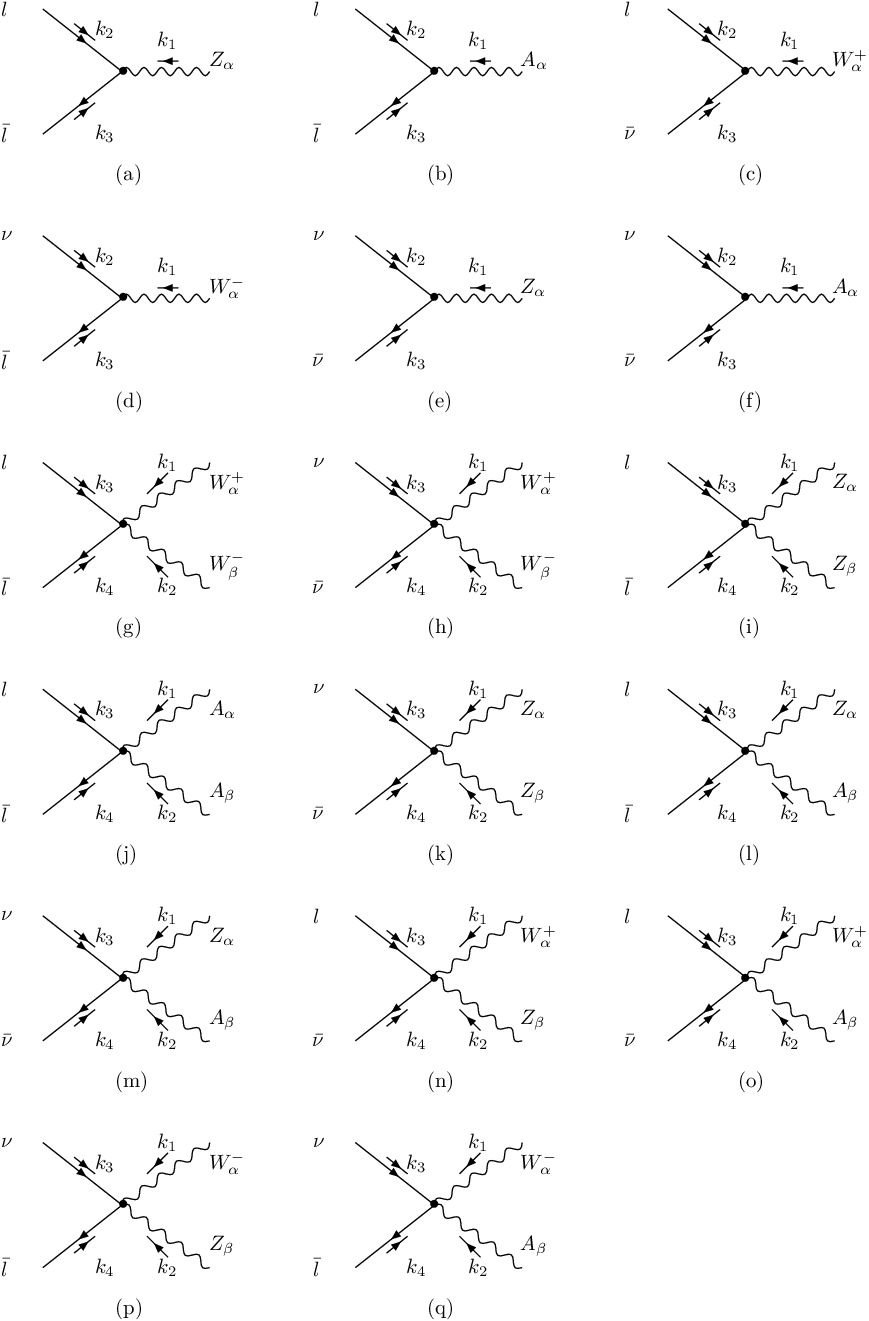}
\label{fig_feynrule}
\end{figure}

\begin{eqnarray}
(a)&&-\frac{1}{\Lambda^2}[k_1{\!\!\!\!\!\slash\;}(k_2-k_3)^\alpha-\gamma^\alpha
k_1\cdot(k_2-k_3)](\frac{c}{2}f_{7}P_L+sf_{11}P_L+sf_{13}P_R) \nonumber \\
&&+\frac{i}{\Lambda^2}[k_1{\!\!\!\!\!\slash\;}k_1^\alpha-k_1^2\gamma^\alpha](\frac{c}{2}f_{24}P_L+sf_{26}P_L+sf_{27}P_R) \nonumber\\
\nonumber\\
(b)&&-\frac{1}{\Lambda^2}[k_1{\!\!\!\!\!\slash\;}(k_2-k_3)^\alpha-\gamma^\alpha
k_1\cdot(k_2-k_3)](\frac{s}{2}f_{7}P_L-cf_{11}P_L-cf_{13}P_R) \nonumber \\
&&+\frac{i}{\Lambda^2}[k_1{\!\!\!\!\!\slash\;}k_1^\alpha-k_1^2\gamma^\alpha](\frac{s}{2}f_{24}P_L-cf_{26}P_L-cf_{27}P_R) \nonumber\\
\nonumber\\
(c)&&\frac{1}{\Lambda^2}[k_1{\!\!\!\!\!\slash\;}(k_2-k_3)^\alpha-\gamma^\alpha
k_1\cdot(k_2-k_3)](\frac{\sqrt{2}}{2}f_{7}P_L)  \nonumber \\
&&+\frac{i}{\Lambda^2}[k_1{\!\!\!\!\!\slash\;}k_1^\alpha-k_1^2\gamma^\alpha](-\frac{\sqrt{2}}{2}f_{24}P_L) \nonumber\\
\nonumber\\
(d)&&\frac{1}{\Lambda^2}[k_1{\!\!\!\!\!\slash\;}(k_2-k_3)^\alpha-\gamma^\alpha
k_1\cdot(k_2-k_3)](\frac{\sqrt{2}}{2}f_{7}P_L)  \nonumber \\
&&+\frac{i}{\Lambda^2}[k_1{\!\!\!\!\!\slash\;}k_1^\alpha-k_1^2\gamma^\alpha](-\frac{\sqrt{2}}{2}f_{24}P_L) \nonumber\\
\nonumber\\
(e)&&-\frac{1}{\Lambda^2}[k_1{\!\!\!\!\!\slash\;}(k_2-k_3)^\alpha-\gamma^\alpha
k_1\cdot(k_2-k_3)](-\frac{c}{2}f_{7}P_L+sf_{11}P_L) \nonumber \\
&&+\frac{i}{\Lambda^2}[k_1{\!\!\!\!\!\slash\;}k_1^\alpha-k_1^2\gamma^\alpha](-\frac{c}{2}f_{24}P_L+sf_{26}P_L) \nonumber\\
\nonumber\\
(f)&&-\frac{1}{\Lambda^2}[k_1{\!\!\!\!\!\slash\;}(k_2-k_3)^\alpha-\gamma^\alpha
k_1\cdot(k_2-k_3)](-\frac{s}{2}f_{7}P_L-cf_{11}P_L) \nonumber \\
&&+\frac{i}{\Lambda^2}[k_1{\!\!\!\!\!\slash\;}k_1^\alpha-k_1^2\gamma^\alpha](-\frac{s}{2}f_{24}P_L-cf_{26}P_L) \nonumber
\end{eqnarray}

\begin{eqnarray}
(g)&&\frac{g}{2\Lambda^2}[g^{\alpha\beta}(k_1{\!\!\!\!\!\slash\;}+k_2{\!\!\!\!\!\slash\;})
+\gamma^\alpha(k_4-k_1-k_3)^\beta+\gamma^\beta(k_3-k_2-k_4)^\alpha](f_{7}P_L) \nonumber \\
&&-\frac{ig}{2\Lambda^2}[g^{\alpha\beta}(k_1{\!\!\!\!\!\slash\;}-k_2{\!\!\!\!\!\slash\;})
+\gamma^\alpha(-2k_1-k_2)^\beta+\gamma^\beta(k_1+2k_2)^\alpha](f_{24}P_L) \nonumber\\
\nonumber\\
(h)&&\frac{g}{2\Lambda^2}[g^{\alpha\beta}(k_1{\!\!\!\!\!\slash\;}+k_2{\!\!\!\!\!\slash\;})
+\gamma^\alpha(k_3-k_1-k_4)^\beta+\gamma^\beta(k_4-k_2-k_3)^\alpha](f_{7}P_L) \nonumber \\
&&+\frac{ig}{2\Lambda^2}[g^{\alpha\beta}(k_1{\!\!\!\!\!\slash\;}-k_2{\!\!\!\!\!\slash\;})
+\gamma^\alpha(-2k_1-k_2)^\beta+\gamma^\beta(k_1+2k_2)^\alpha](f_{24}P_L) \nonumber\\
\nonumber\\
(i)&&-\frac{g}{\Lambda^2}[g^{\alpha\beta}(k_1{\!\!\!\!\!\slash\;}+k_2{\!\!\!\!\!\slash\;})
-\gamma^\alpha k_1^\beta-\gamma^\beta
k_2^\alpha](\frac{2s^2-1}{2}f_{7}P_L+\frac{s(2s^2-1)}{c}f_{11}P_L+\frac{2s^3}{c}f_{13}P_R) \nonumber\\
\nonumber\\
(j)&&\frac{g}{\Lambda^2}[g^{\alpha\beta}(k_1{\!\!\!\!\!\slash\;}+k_2{\!\!\!\!\!\slash\;})
-\gamma^\alpha k_1^\beta-\gamma^\beta
k_2^\alpha](s^2f_{7}P_L-2csf_{11}P_L-2csf_{13}P_R) \nonumber\\
\nonumber\\
(k)&&\frac{g}{\Lambda^2}[g^{\alpha\beta}(k_1{\!\!\!\!\!\slash\;}+k_2{\!\!\!\!\!\slash\;})
-\gamma^\alpha k_1^\beta-\gamma^\beta
k_2^\alpha](\frac{1}{2}f_{7}P_L-\frac{s}{c}f_{11}P_L) \nonumber
\end{eqnarray}

\begin{eqnarray}
(l)&&\frac{g}{\Lambda^2}[g^{\alpha\beta}k_1{\!\!\!\!\!\slash\;}
-\gamma^\alpha k_1^\beta](csf_{7}P_L+2s^2f_{11}P_L) \nonumber\\
&&+\frac{g}{\Lambda^2}[g^{\alpha\beta}k_2{\!\!\!\!\!\slash\;}
-\gamma^\beta k_2^\alpha](-\frac{s(2s^2-1)}{2c}f_{7}P_L+(2s^2-1)f_{11}P_L  \nonumber\\
\nonumber\\
(m)&&\frac{g}{\Lambda^2}[g^{\alpha\beta}k_2{\!\!\!\!\!\slash\;}
-\gamma^\beta k_2^\alpha](\frac{s}{2c}f_{7}P_L+f_{11}P_L)  \nonumber\\
\nonumber\\
(n)&&-\frac{g}{\Lambda^2}\frac{\sqrt{2}}{2}c[\gamma^\alpha(k_3-k_4)^\beta-\gamma^\beta(k_3-k_4)^\alpha]f_{7}P_L \nonumber\\
&&+\frac{g}{\Lambda^2}\frac{\sqrt{2}s^2}{2c}[g^{\alpha\beta}k_1{\!\!\!\!\!\slash\;}-\gamma^\alpha
k_1^\beta]f_{7}P_L \nonumber\\
&& -\frac{g}{\Lambda^2}\sqrt{2}s[g^{\alpha\beta}k_2{\!\!\!\!\!\slash\;}-\gamma^\beta
k_2^\alpha]f_{11}P_L \nonumber\\
&& -\frac{ig}{\Lambda^2}\frac{\sqrt{2}c}{2}[g^{\alpha\beta}(k_1{\!\!\!\!\!\slash\;}-k_2{\!\!\!\!\!\slash\;})-\gamma^\alpha
(2k_1+k_2)^\beta+\gamma^\beta (k_1+2k_2)^\alpha]f_{24}P_L \nonumber\\
\nonumber\\
(o)&&-\frac{g}{\Lambda^2}\frac{\sqrt{2}}{2}s[\gamma^\alpha(k_3-k_4)^\beta-\gamma^\beta(k_3-k_4)^\alpha]f_{7}P_L \nonumber\\
&&-\frac{g}{\Lambda^2}\frac{\sqrt{2}s}{2}[g^{\alpha\beta}k_1{\!\!\!\!\!\slash\;}-\gamma^\alpha
k_1^\beta]f_{7}P_L \nonumber\\
&& +\frac{g}{\Lambda^2}\sqrt{2}c[g^{\alpha\beta}k_2{\!\!\!\!\!\slash\;}-\gamma^\beta
k_2^\alpha]f_{11}P_L \nonumber\\
&& -\frac{ig}{\Lambda^2}\frac{\sqrt{2}s}{2}[g^{\alpha\beta}(k_1{\!\!\!\!\!\slash\;}-k_2{\!\!\!\!\!\slash\;})-\gamma^\alpha
(2k_1+k_2)^\beta+\gamma^\beta (k_1+2k_2)^\alpha]f_{24}P_L \nonumber\\
\nonumber\\
(p)&&\frac{g}{\Lambda^2}\frac{\sqrt{2}}{2}c[\gamma^\alpha(k_3-k_4)^\beta-\gamma^\beta(k_3-k_4)^\alpha]f_{7}P_L \nonumber\\
&&+\frac{g}{\Lambda^2}\frac{\sqrt{2}s^2}{2c}[g^{\alpha\beta}k_1{\!\!\!\!\!\slash\;}-\gamma^\alpha
k_1^\beta]f_{7}P_L \nonumber\\
&&-\frac{g}{\Lambda^2}\sqrt{2}s[g^{\alpha\beta}k_2{\!\!\!\!\!\slash\;}-\gamma^\beta
k_2^\alpha]f_{11}P_L \nonumber\\
&& +\frac{ig}{\Lambda^2}\frac{\sqrt{2}c}{2}[g^{\alpha\beta}(k_1{\!\!\!\!\!\slash\;}-k_2{\!\!\!\!\!\slash\;})-\gamma^\alpha
(2k_1+k_2)^\beta+\gamma^\beta (k_1+2k_2)^\alpha]f_{24}P_L \nonumber\\
\nonumber\\
(q)&&\frac{g}{\Lambda^2}\frac{\sqrt{2}}{2}s[\gamma^\alpha(k_3-k_4)^\beta-\gamma^\beta(k_3-k_4)^\alpha]f_{7}P_L \nonumber\\
&&-\frac{g}{\Lambda^2}\frac{\sqrt{2}s}{2}[g^{\alpha\beta}k_1{\!\!\!\!\!\slash\;}-\gamma^\alpha
k_1^\beta]f_{7}P_L \nonumber\\
&&+\frac{g}{\Lambda^2}\sqrt{2}c[g^{\alpha\beta}k_2{\!\!\!\!\!\slash\;}-\gamma^\beta
k_2^\alpha]f_{11}P_L \nonumber\\
&& +\frac{ig}{\Lambda^2}\frac{\sqrt{2}s}{2}[g^{\alpha\beta}(k_1{\!\!\!\!\!\slash\;}-k_2{\!\!\!\!\!\slash\;})-\gamma^\alpha
(2k_1+k_2)^\beta+\gamma^\beta (k_1+2k_2)^\alpha]f_{24}P_L \nonumber
\end{eqnarray}

\end{document}